\documentclass[submission,copyright,creativecommons]{eptcs}
\providecommand{\event}{MBT 2012} 

\usepackage{savesym}
\savesymbol{hbar}
\usepackage{semantic}
\usepackage{czt}
\usepackage{multicol}
\usepackage{lscape}
\usepackage{tabularx}

\reservestyle{\command}{\mathsf}
\command{define,->,bool,lambda,and,or,tuple,subtype,select,%
         false,true,record,assert,exists,forall,if,nat,int,%
         define-type,scalar,nat1,NAT,INT,SUBTYPE,LAMBDA,TYPE,%
         DATATYPE,END,ARRAY,OF,BITVECTOR,BV1,0,IF,THEN,ELSE,%
         ENDIF,AND,ASSERT,FORALL,1,BOOLEAN,NAT1}

\title{Applying SMT Solvers to the Test Template Framework}
\author{Maximiliano Cristi\'a
\institute{CIFASIS and UNR\\ Rosario, Argentina}
\email{cristia@cifasis-conicet.gov.ar}
\and
Claudia Frydman
\institute{LSIS-CIFASIS\\
Marseille, France}
\email{claudia.frydman@lsis.org}
}
\def\titlerunning{Applying SMT Solvers to the TTF}
\def\authorrunning{M. Cristi\'a, \& C. Frydman}

\begin{document}
\maketitle

\begin{abstract}
The Test Template Framework (TTF) is a model-based testing method for the Z notation. In the TTF, test cases are generated from test specifications, which are predicates written in Z. In turn, the Z notation is based on first-order logic with equality and Zermelo-Fraenkel set theory. In this way, a test case is a witness satisfying a formula in that theory. Satisfiability Modulo Theory (SMT) solvers are software tools that decide the satisfiability of arbitrary formulas in a large number of built-in logical theories and their combination. In this paper, we present the first results of applying two SMT solvers, Yices and CVC3, as the engines to find test cases from TTF's test specifications. In doing so, shallow embeddings of a significant portion of the Z notation into the input languages of Yices and CVC3 are provided, given that they do not directly support Zermelo-Fraenkel set theory as defined in Z. Finally, the results of applying these embeddings to a number of test specifications of eight cases studies are analysed.
\end{abstract}

\section{Introduction}
The Test Template Framework (TTF) is a model-based testing (MBT) method for the Z notation, specially well-suited for unit testing \cite{Stocks2}. The Z notation is a formal specification language based on first-order logic with equality and Zermelo-Fraenkel set theory \cite{Spivey00, ZISO}. Our group was the first in providing tool support for the TTF by implementing Fastest \cite{CristiaPRM, CristiaARM}, and in extending the TTF beyond test case generation \cite{CristiaPluss, CristiaFTCRL}. 

Within the TTF, each operation of a Z specification is analysed to produce test cases to later test its implementation. This analysis is performed by partitioning the input space of the operation. Partitioning, in turn, is conducted through the application of one or more testing tactics. Each element of the resulting partition is an equivalence class. In this context, equivalence classes are called \textit{test classes}, {\it test objectives}, {\it test templates} or {\it test specifications} in the literature. The latter will be used in this paper. Test specifications obtained in this way can be further subdivided into more test specifications by applying other testing tactics. The net effect of this technique is a progressive partition of the input space into more restrictive test specifications. One of the features that makes the TTF particularly appealing for the Z community, is that all of its main concepts are expressed in Z. Hence, the engineer has to know only one notation to perform the analysis down to the generation of abstract test cases.

Each test specification is characterized by a Z predicate. Finding a test case for a test specification in the TTF means, thus, finding a witness satisfying its predicate. Clearly, this is a problem of satisfiability at the presence of a complex and rich mathematical theory. Currently, Fastest implements a rough algorithm to solve this problem \cite{CristiaPRM}. On the other hand, Satisfiability Modulo Theory (SMT) solvers are tools that, precisely, solve the problem of satisfiability for a range of mathematical and logical theories \cite{Nieuwenhuis00}. In this paper, we present the first results of applying two SMT solvers, Yices \cite{Dutertre00} and CVC3 \cite{Barrett00}, to the problem of finding test cases from test specifications within the TTF.

Applying a SMT solver to this problem is not a trivial task, in part, due to the fact that, as far as we have investigated, currently there is no SMT solver natively supporting the Zermelo-Fraenkel set theory. Hence, one needs to rest on defining a shallow embedding of that theory in the language of a SMT solver. In doing so, a key question arises: is the language of a SMT solver expressive enough to allow an embedding of Zermelo-Fraenkel set theory? Then, if that embedding is possible, is it the only one? Will the chosen embedding solve all the satisfiable test specifications appearing in the TTF and real Z specifications? Which SMT solver and which embedding will be the best in satisfying more test specifications in less time? Finally, some questions more specific to our project: will that SMT solver be better than Fastest in finding test cases? Or should the SMT solver complement Fastest in this task?

In this paper we give first answers to all these questions. In Section \ref{z} we describe some issues about the Z notation that pose some requirements on the expressiveness of SMT solvers' languages. Section \ref{ttf} shows the complexity of typical test specifications derived by applying the TTF, and Section \ref{genalltca} briefly describes the algorithm implemented in Fastest to search for test cases. A research plan for the application of SMT solvers to this problem is established in Section \ref{plan}. Sections \ref{embeddings} and \ref{experiments} present the embeddings for Yices and CVC3 and the results of an empirical assessment of them, respectively. Finally, in Sections \ref{relwork} and \ref{concl} we compare our work with other approaches and give our conclusions.

\section{\label{z}The Z Notation}
In this section we do not pretend to introduce the Z notation but only to highlight some peculiarities of its type system---for a comprehensive presentation of Z there are many fine textbooks \cite{Potter, Jacky}. An important component of the Z notation is the Z Mathematical Toolkit (ZMT) \cite{Spivey00}. The ZMT defines a number of mathematical data structures and operations on them. It contains all the elements of the Zermelo-Fraenkel set theory and other elements built on them. In this context, we will refer to the ZMT as a synonym of first-order logic with equality and Zermelo-Fraenkel set theory.

Z is a typed formalism. $\num$ is the only built-in type in the language. Specifiers can introduce basic types as they wish by simply declaring them as: $[X]$. The structure of the elements of such a type are unknown. It is also possible to introduce so-called free types, which are recursive data types. In their simplest form they are just enumerations: $Y ::= y_1 | \dots | y_n$.

Basically, Z has three type constructors. If $X$ is a type, then $\power X$ builds the type of all the sets whose elements are of type $X$. 
If $X$ and $Y$ are types then, $X \cross Y$ is the type of ordered pairs or Cartesian product\footnote{In Z an ordered pair is usually written as $x \mapsto y$ as a synonym of $(x,y)$.}. Finally, if $X_1, \dots, X_k$ are $k$ types, then $[x_1:X_1; \dots; x_k:X_k]$ is the type of records whose fields are $x_1, \dots, x_k$. In Z records are called schema types, or just schemas, and are central to the notation---they are used to specify states, operations, properties, etc. These type constructors can be applied recursively to form more and more complex types. 

The ZMT also defines a number of synonyms for some important sets. The set of all binary relations between $X$ and $Y$, noted as $X \rel Y$, is defined as $\power (X \cross Y)$. 
Furthermore, the ZMT next defines the set of partial functions from $X$ to $Y$, $X \pfun Y$, and the set of total functions, $X \fun Y$, as:
\[
X \pfun Y == 
  \{f:X \rel Y | 
      \forall x:X; y_1,y_2:Y @ 
         x \mapsto y_1 \in f \land x \mapsto y_2 \in f \implies y_1 = y_2\} \also
X \fun Y == \{f: X \pfun Y | \dom f = X\}
\]

That is, in Z functions are sets of ordered pairs. In other words, functions are built up from more basic elements, are not always total, they are extensional, and can be higher-order---i.e. a function can have another function as an argument. Being sets of ordered pairs, set theory operators can be applied to them: if $f:X \pfun Y; x:X; y:Y$ then the following are all type correct $x \mapsto y \in f$, $f \cup \{x \mapsto y\}$, $f \setminus \{x \mapsto y\}$, etc. However, they are also functions so function application is also defined: $f~x = f(x)$. Therefore, Z functions have two characters: they are functions but they are also sets. The Z type system cannot guarantee that, for instance $f \cup \{x \mapsto y\}$, is still a partial function, it can only guarantee that it is a binary relation. Moreover, $f~x$ is not always defined since it might be the case that $x \notin \dom f$. That is, Z cannot guarantee that function application is always correct. All this is crucial to the complexity of embedding Z in other languages because usually functions are types in their own and are total, like in Yices \cite{Dutertre00}, sometimes they are just first-order objects like in Z3 \cite{Bjorner00} and they may be non-extensional and total like in CVC3 \cite{Barrett00} and Coq \cite{CoqRM}. 

$\power$ builds all the sets of a given type, both finite and infinite. Therefore, the ZMT defines the set of finite sets of a type:
\[
\finset X == \{S: \power X | \exists n:\nat @ \exists f:1 \upto n \fun S @ \ran f = S\}
\]

\noindent over which the cardinality operator, $\#$, can be applied. That is, $\#$ cannot be applied to $A:\power X$ unless you can prove that $A$ is actually in $\finset X$. The ZMT also defines the sets of finite partial functions and sequences:
\[
X \ffun Y == \{f:X \pfun Y | \dom f \in \finset X\} \also
\seq X == \{f: \nat \ffun X | \dom f = 1 \upto \#f\}
\]

The last issue we want to remark about the Z notation is that set theory operators are polymorphic. In other words, symbols like $\cup$, $\cap$, $\in$ and $\emptyset$, can be applied to any type.




\section{\label{ttf}Test Specifications and Test Cases in the TTF}
Test specifications and test cases in the TTF are represented as Z schemas. In its more complex form a Z schema has two parts: the declaration part, where variables are declared; and the predicate part, where a predicate depending on that variables can be written. The next schemas are test specifications borrowed from two of our case studies:

{\small
\begin{multicols}{2}
\setlength{\columnsep}{18.0pt}
\begin{schema}{DetectReferenceEvent^{NR}_{18}}\\
 now, fa : \nat;
 ot : REVENT \pfun \nat \\
 tli, tls, X: REVENT \fun \nat \\
 sysState : STATUS; e? : REVENT 
\where
 e? = LiftOff \\
 sysState = normal \\
 e? \notin \dom ot \\
 now \in tli~e? \upto tls~e? \\
 X~e? \leq fa \\
 ot \neq \emptyset \\
 \{ e? \mapsto now \} \neq \emptyset \\
 ot \cap \{ e? \mapsto now \} = \emptyset \\
 1 < now < 3
\end{schema}

\columnbreak

\begin{schema}{RetrieveEData^{SP}_{24}}\\
 mem: \seq MDATA \\
 m: \nat \\
 d? : \seq MDATA 
\where
 43 < m + \# d? \\
 mem \neq \emptyset \\
 \{ i : 1 \upto \# d? @ m + i \mapsto d?~i \} \neq \emptyset \\
 \dom mem \cap \dom \{ i : 1 \upto \# d? @ m + i \mapsto d?~i \} \neq \emptyset \\
 \lnot \dom \{ i : 1 \upto \# d? @ m + i \mapsto d?~i \} \subseteq \dom mem \\
 \lnot \dom mem \subseteq \dom \{ i : 1 \upto \# d? @ m + i \mapsto d?~i \}
\end{schema}
\end{multicols}
}

As it can be seen, test specifications are conjunctions of atomic predicates making heavy use of sets, functions, sequences and set theory operators. A test case is, then, a schema further restricting its test specification so the declared variables can take only one value. For example, the following schema represents a test case generated from $DetectReferenceEvent^{NR}_{18}$. 
\vspace{-0.5cm}

\begin{schema}{DetectReferenceEvent^{TC}_{18}}\\
DetectReferenceEvent^{NR}_{18}
\where
 tli = 
  \{LiftOff \mapsto 2, ThrustDrop1E \mapsto 5, ThrustDrop2E \mapsto 4, ThrustDrop3E \mapsto 10\} \\
 tls = 
  \{LiftOff \mapsto 10, 
     ThrustDrop1E \mapsto 12, ThrustDrop2E \mapsto 14, ThrustDrop3E \mapsto 16\} \\
 X = 
  \{LiftOff \mapsto 3, ThrustDrop1E \mapsto 5, ThrustDrop2E \mapsto 7, ThrustDrop3E \mapsto 9\} \\
 e? = LiftOff \\
 sysState = normal \\
 now = 2 \\
 fa = 10 \\
 ot = \{ThrustDrop1E \mapsto 3\} 
\end{schema}

Note how schema inclusion is used to link a test case with its corresponding test specification.

\section{\label{genalltca}A Simple Algorithm for Searching Test Cases}
Before searching test cases from test specifications, Fastest's users can run a command to eliminate unsatisfiable test specifications. The method behind this command has been extensively described elsewhere \cite{CristiaARM}. This method has proved to be efficient and effective in eliminating most of the unsatisfiable test specifications. Hence, when users want to find test cases from test specifications, most of them are satisfiable. Fastest implements a very simple algorithm to search test cases from test specifications, which has been introduced in another paper \cite{CristiaPRM}. At the time we started Fastest (early 2007) SMT solvers were not an option since most of them were being developed at the same time. Then, we implemented a primitive algorithm that can be regarded as a ``brute force ZMT solver''\footnote{The `Z' in `ZMT solver' is not a mistake, but an indication that our algorithm is only for the Z Mathematical Toolkit.}. Fastest builds a finite model for each test specification by calculating the Cartesian product between a very small finite set of values bound to each variable declared in the test specification. Later, Fastest evaluates the test specification for some elements in the finite model. These finite models are calculated by considering the following heuristics:

\begin{itemize}\label{fm}
\item Only the types of variables are considered when building the finite model; i.e. the structure of the predicate appearing in the test specification is not taken into account.

\item There is a configuration variable, $FSS$, whose value sets the size of the finite sets for basic types, $\num$ and $\nat$. $FSS$ must be strictly positive---usually it is 2 or 3.

\item There is a configuration variable, $MAX$, whose value sets the maximum size for a finite model.

\item The finite sets for types $\nat$ or $\num$ are built from the first $FSS$ numerical constants appearing in the test specification. If there are no such constants then $[0 \upto FSS - 1]$ is chosen for $\nat$ and $[-(FSS \div 2 + (FSS \mod 2 - 1)) \upto (FSS \div 2)]$ for $\num$.

\item The finite sets for enumerated types are their elements.

\item The finite sets for basic types are built by generating $FSS$ constant names of each type.

\item If a variable declared in the test specification does not appear in its predicate, then the finite set for that variable is any singleton---since the value of such a variable has no influence whatsoever on the evaluation of the predicate.

\item If the predicate of a test specification contains an atomic predicate of the form $var = val$, where $var$ is a variable and $val$ is a constant value, then the finite set for $var$ is just $\{val\}$---since it will be impossible to satisfy the predicate with any other value.

\item The finite sets for the types or sets that result from applying a type constructor or by following a ZMT definition---i.e. $\cross$, $\power$, $\pfun$, etc.---to other types or sets, are built recursively from the finite sets considered for its arguments.

\item Given that test specifications are conjunctions of atomic predicates, the algorithm reduces the initial finite model to the subset satisfying the first atomic predicate. This subset is used as the finite model on which the second atomic predicate is evaluated, and the algorithm reduces it once more to the subset satisfying this second predicate. This continues until the last atomic predicate is considered, in which case the first element satisfying it is returned; or until an atomic predicate cannot be satisfied, in which case ``unknown'' is returned. 

During this step: (a) $MAX$ is considered to put a limit on the number of elements of the finite model to be explored; and (b) the evaluation of the predicate on a particular element of the finite model is performed by the ZLive component of the CZT project \cite{CZT, Malik00}.
\end{itemize}

Although this algorithm might appear inefficient and is certainly inelegant, it has proved to find an average of 80\% of test cases from satisfiable test specifications \cite{CristiaPRM}. However, SMT solvers can be good complements or alternatives to this algorithm, as we will show shortly.

\section{\label{plan}Contribution of this Paper}
Replacing the algorithm described in the previous section, is not a trivial task because, just to begin with, no SMT solver works directly with the Z notation. Therefore, at a bare least we need to write a translator from Z into the language of the chosen SMT solver, and another from the output language back to Z---for converting the witness found by the SMT solver into Z---, when even the subset of Z supported by Fastest is a complex language. Not to mention that it might be necessary to try out different SMT solvers with different shallow embeddings. Hence, we plan to attack this problem as follows:

\begin{enumerate}
\item\label{e:chose} Chose some SMT solvers that are powerful and stable as to be used for the problem at hand.
\item Define one or more shallow embeddings for them.
\item Apply the embeddings to the satisfiable test specifications that were not solved by Fastest.
\item\label{e:analyse} Analyse the results.
\item If the combination of SMT solver and shallow embedding works well for these test specifications---i.e., it finds many test cases fast---, then see whether it also finds test cases for those test specifications for which Fastest works well.
\item Since ZLive has some limitations, see if the shallow embedding can overcome them.
\item Complete and optimise the embedding.
\item\label{e:trad} If everything goes well, write the traducers.
\item Measure the end-to-end computing time---i.e., translating from Z to the SMT solver, executing the SMT solver, and translating back the results to Z---to compare it with the current algorithm.
\item Investigate whether the SMT solver can be used to eliminate unsatisfiable test specifications.
\end{enumerate}

Until step \ref{e:trad} all the work is manual and many alternatives should be constantly evaluated. For example, is a single SMT solver good enough? Is better to use many of them because some solve some test specifications while others solve the rest? Must the current algorithm be replaced or used as another solver? At the end, would it be better to write a decision procedure for the ZMT and include it in some SMT solver instead of using a shallow embedding?

In this paper we address steps \ref{e:chose} to \ref{e:analyse}. More specifically, the problem attacked in this paper is, thus, using SMT solvers to find witnesses satisfying those test specifications for which Fastest failed---two of which are shown in Section \ref{ttf}. 
Our contributions are: (i) defining shallow embeddings of a significant portion of the ZMT for two mainstream SMT solvers, namely Yices and CVC3; and (ii) running Yices and CVC3 on 69 satisfiable test specifications (borrowed from eight cases studies, three of which are real industrial problems) written with the shallow embeddings, to measure the efficiency and effectiveness of the embeddings and the SMT solvers for this particular testing problem. The embeddings shown in this paper are not only useful for our problem but also for others as they embed the Zermelo-Fraenkel set theory in general \cite{Kroning00}.

\section{\label{embeddings}Shallow Embeddings of Z into Two SMT Solvers}
In this section we present two shallow embeddings of a significant portion of the ZMT for Yices \cite{Dutertre00} and CVC3 \cite{Barrett00}---in Section \ref{relwork} we explain why we have chosen these two SMT solvers. The embeddings are given by means of embedding rules of the following form:
\[
\inference[name]{Z\ notation}{\mathsf{SMT\ solver\ syntax}}
\]
\noindent where the text above the line is some Z term and the text below the line is one or more,  either Yices or CVC3, sentences; the name of the rule identifies the Z term being considered. 
Some Z features are omitted because they are outside the scope of this paper; and some rules are not given because they can be easily deduced from the others (for example, we give a rule for set intersection but not for set union). 

The files resulting from applying these embeddings to 69 test specifications along with the Z test specifications themselves are available at: \verb+www.fceia.unr.edu.ar/~mcristia/smt-ttf-cs.tar.gz+.

\subsection{Notation}
We decided to describe the embeddings in terms of the input languages of Yices and CVC3 
because we would like readers to be able to check all the empirical data mentioned in Section \ref{experiments}. We do not use the SMT-LIBv2 \cite{Cok00} language because it does not support all the features of all SMT solvers---for instance, Yices' lambda expressions and CVC3's instantiation patters. We believe that, in general, the input languages of both SMT solvers are rather easy to understand for readers knowledgeable in formal methods. 

Yices uses a language similar to SMT-LIBv2. That is, operators and type constructors are all prefix. For instance, $x + y$ is written $(+~x~y)$, and a function from $X$ to $Y$ is declared as $(-> X~Y)$. $\<nat>$, $\<int>$ and $\<bool>$ are all built-in types, with their obvious meanings. Yices support lambda expressions to define functions, as in lambda calculus. The keyword $\<select>$ is used to access members of record-types; it is also a prefix operator. 

CVC3 uses a more human-readable input language. All the reserved words are written in capitals. The most difficult construction is the definition of an array. If $A$ is an array then it is possible to associate a value for each of its components by means of the construction $\<ARRAY>~(x:T):~expr(x)$, where $T$ is the type of the indexes of $A$, and $expr$ is an expression of the type of the components of $A$ which may depend on $x$. The result is an array in which the value of the component with index $x$ is the result of $expr(x)$.


\subsection{Yices}
The most relevant rules of the shallow embedding of Z into Yices\footnote{Actually we used Yices 1.} are given in Figure \ref{yices1}. We are going to discuss only those rules that deserve some attention. As it can be seen sets, functions, partial functions, binary relations, sequences and finite sets are all represented, essentially, with Yices uninterpreted functions. In Yices uninterpreted functions\footnote{From now on we will just say ``functions''.} are total, extensional and higher-order, making them a good choice to represent ZMT's mathematical structures. In our opinion, there are no other elements in Yices better than functions on which to build the embedding. 

\begin{figure}
\begin{array}[t]{ll}
\inference[$\num$]{\num}{\<int>} \t5
\inference[$\nat$]{\nat}{\<nat>} &
\inference[basic types]{[X]}{(\<define-type>~X)} \\\noalign{\smallskip}
\inference[free types]{X ::= c_1 | \dots | c_n}{(\<define-type>~X~(\<scalar>~c_1 \dots c_n))} &
\inference[$\cross$]{x: Y \cross Z}{(\<define>~x::[Y,Z])} \\\noalign{\smallskip}
\inference[$\power$]{A: \power X}{(\<define>~A::(-> X~\<bool>))} &
\inference[ranges]{a \upto b}{(\<lambda>~(i::\<int>)~(\<and>~(\leq a~i)~(\leq i~b)))} \\\noalign{\smallskip}
\inference[$\rel$]{R: X \rel Y}{(\<define>~R::(-> X~Y~\<bool>))} &
\inference[$\fun$]{f: X \fun Y}{(\<define>~f::(-> X~Y))} \\\noalign{\smallskip}
\end{array}
\begin{array}[t]{l}
\inference[$\pfun$]{f: X \pfun Y}{(\<define>~ f::(\<record>~ dom::(-> X~\<bool>)~ law::(-> X~Y)))} \\\noalign{\smallskip}
\inference[$\emptyset$]{\emptyset: X}{(\<define>~ emptysetX::(-> X~\<bool>)~(\<lambda>~ (x::X)~\<false>))} \\\noalign{\smallskip}
\inference[$\cap$]{A,B: \power X & A \cap B}
  {\begin{array}[t]{l}
     (\<define>~capX::(-> (-> X~\<bool>)~(-> X~\<bool>)~(-> X~\<bool>)) \\
       \t1 (\<lambda>~(A::(-> X~\<bool>)~ B::(-> X~\<bool>))~ (\<lambda>~(x::X)~(\<and>~(A~x)~(B~x)))))
   \end{array}
  } \\\noalign{\smallskip}
\inference[$\subseteq$]{A,B: \power X & A \subseteq B}
  {\begin{array}[t]{l}
    (\<define>~subseteqX::(-> (-> X~\<bool>)~(-> X~\<bool>)~\<bool>) \\
      \t1 (\<lambda>~(A::(-> X~\<bool>)~B::(-> X~\<bool>))~(\<forall>~(x::X)~(=> (A~x)~(B~x)))))
   \end{array}
  }  \\\noalign{\smallskip}
\inference[$\finset$]{A:\finset X}
  {\begin{array}[t]{l}
    (\<define>~A::(\<record>~set::(-> X~\<bool>)~bij::(-> X~\<nat1>)~card::\<nat>)) \\
    (\<assert>~(\<forall>~(x::X)~
         (<=> ((\<select>~A~set)~x)~(\leq ((\<select>~A~bij)~x)~(\<select>~A~card))))) \\
    (\<assert>~(\<forall>~(n::\<nat1>~x_1::X~x_2::X)\\
      \t1 (=> (\<and>~(\leq n~(\<select>~A~card)) \\
          \t3 ((\<select>~A~set)~x_1) \\
          \t3 ((\<select>~A~set)~x_2) \\
          \t3 (= ((\<select>~A~bij)~x_1)~n) \\
          \t3 (= ((\<select>~A~bij)~x_2)~n)) \\
      \t2 (= x_1~x_2))))
   \end{array}
  } \\\noalign{\smallskip}
\inference[$\seq$]{s:\seq X}
  {\begin{array}[t]{l}
    (\<define>~s::(\<record>~dom::(-> \<nat1>~\<bool>)~law::(-> \<nat1>~X)~card::\<nat>)) \\
    (\<assert>~(\<forall>~(n::\<nat1>)~(<=> (\leq n~(\<select>~s~card))~((\<select>~s~dom)~n))))
   \end{array}
  }
\end{array}
\caption{\label{yices1}Embedding rules for Yices.}
\end{figure}

Basic types are embedded as type definitions thus preserving the Z semantics in that there is no clue about the structure of their elements. The embedding defines a set of type $X$ as a function from $X$ to $\<bool>$. If $A:\power X$ and $x:X$, the interpretation is trivial: $x \in A \iff (A~x)$. Note that these two rules imply that a set may be infinite. Also, note that Yices' type system impedes us to define polymorphic operators (cf. rules $\emptyset$, $\subseteq$, etc.). The workaround is to define one per type. This is not a serious problem because the intention is that the embedding will be transparent for Fastest users.

A partial function is represented as a record with two fields: $dom$, is a Yices function representing the domain of the function---as with sets---; and $law$ is the actual map between the types. The intention is that $(law~x)$ is meaningful if and only if $(dom~x)$ is true. However, this intention cannot be guaranteed unless the Z specification is consistent---and not only type correct. If the Z specification is verified in a system like, for instance, Z/EVES \cite{ZEVES}, then some proof obligations should have been discharged proving that all partial functions are correctly applied. Besides, Fastest eliminates test specifications where a partial function is explicitly applied outside its domain---i.e., for example where $x \notin \dom f \land \dots f~x \dots$ holds. Our embedding assumes these two hypothesis. This representation of partial functions has two advantages: (i) the domain is a set as in Z; and (ii) it is easy to apply a function to its argument. However, it has a disadvantage: (partial) functions are not sets; in other words, the embedding for (partial) functions is semantically different from the embedding for sets. For instance, if we have $f:X \pfun Y$ and $R:X \rel Y$, then at the Z level $f = R$ is type-correct, but its representation through the embedding is not. Nevertheless, this can be overcome by calculating the ``set'' corresponding to a (partial) function. For example (assuming $f:X \pfun Y$):
\[
(\<define>~fSet::(-> X~Y~bool)\\
  \t1 (\<lambda>~(x::X~y::Y)~(\<and>~((\<select>~f~dom)~x)~(= ((\<select>~f~law)~x)~y))))
\]

\noindent Then, at the Yices level we can write $(= fSet~R)$ for $f = R$, but we still use $f$ for function application; for instance, $((\<select>~f~law)~x = y_1)$. We have tried other representations for partial functions, sets and functions but in our opinion this is the best one for our purposes. For instance, the Yices' manual suggests representing partial functions through dependent types:
\[
(\<define>~f::(\<tuple>~dom::(-> X~bool)~(-> (\<subtype>~(x::X)~(dom~x))~Y)))
\]

\noindent However, it has a problem in the context of embedding Z specifications. In effect, if $x:X$ then $f~x$ is type-correct in Z, but $((\<select>~f~2)~x)$ is not in Yices---because the type of $x$ is $X$ and not $(\<subtype>~ (x::X)~(dom~ x))$. This representation may be good for other theories of partial functions.

As it can be seen, finite sets are harder to represent. We embed them by representing the definition of finiteness given in Section \ref{z}---i.e., a set has cardinality $n$ if there is a bijection between itself and the first $n$ natural numbers. That is, a finite set is a record with three fields: $set$, is the actual set; $bij$, is intended to be a bijection from a subset of its domain onto a subset of its range; and $card$, is the cardinality of the set. To keep $set$ finite and consistent with the other two fields we assert two axioms. The first one says that an element is in $set$ if and only if $bij~x$ is less than or equal to $card$. This ensures that the image of $bij$ for those $x:X$ such that $set~x$ is true, has a finite number of elements. The second axiom asserts that the inverse of $bij$ for all the natural numbers less than or equal to $card$, is a function. Therefore, $bij$ is a bijection between the range $[1 \upto card]$ and all $x:X$ such that $set~x$ is true. Observe, that this representation is compatible with the one for sets. In effect, if $A:\power X; B:\finset X$ and $B \subseteq A$, then at the Yices level we can simply say $(subseteqX~(\<select>~B~set)~A)$ (cf. rule $\subseteq$).

The rules in Figure \ref{yices1} are completed by a rule saying that each Z atomic predicate appearing in a test specification must be embedded as an $\<assert>$ command. This is justified because a test specification is a conjunction of atomic predicates and a sequence of $\<assert>$ commands is also a conjunction. Therefore, checking the satisfiability of a test specification is performed by executing a $\mathsf{check}$ command.

\subsection{CVC3}
The most relevant rules of the shallow embedding of Z into CVC3 are given in Figure \ref{cvc3}. Due to space restrictions we write $\<BV1>$ for $\mathsf{BITVECTOR}(1)$, $\<0>$ for $\mathsf{0bin0}$ and $\<1>$ for $\mathsf{0bin1}$. As it can be seen, the embedding is essentially the same to the previous one, the main difference being that it uses arrays instead of functions. Although CVC3 supports functions, they are not extensional nor higher-order making them less useful to represent the ZMT. On the other hand, CVC3 provides a general theory of extensional, higher-order arrays. In particular they can be indexed by any type, finite or infinite. Therefore, in this case we opted for arrays as the main mathematical structure for the embedding. For sets and the like we used arrays of bit vectors of size one, because in CVC3 arrays cannot have Boolean components. This makes the embedding more verbose than the one for Yices. Note, however, that we have used functions for defining set theory operators like intersection and subset. We believe that this embedding deserves no further comments due to its similarities with respect to the previous one.

\begin{figure}
\begin{array}[t]{ll}
\inference[$\num$]{\num}{\<INT>} &
\inference[basic types]{[X]}{X:\<TYPE>} \\\noalign{\smallskip}
\inference[free types]{X ::= c_1 | \dots | c_n}{\<DATATYPE>~X = c_1 | \dots | c_n~\<END>} &
\inference[$\cross$]{x: Y \cross Z}{x:[Y,Z]} \\\noalign{\smallskip}
\inference[$\power$]{A: \power X}{A:\<ARRAY>~X~\<OF>~\<BV1>} &
\inference[ranges]{a \upto b}{[a..b]} \\\noalign{\smallskip}
\inference[$\rel$]{R: X \rel Y}{A:\<ARRAY>~[X,Y]~\<OF>~\<BV1>} &
\inference[$\fun$]{f: X \fun Y}{A:\<ARRAY>~X~\<OF>~Y}
\end{array}
\begin{array}[t]{l}
\inference[$\nat$]{\nat}{\<NAT>:\<TYPE> = \<SUBTYPE>(\<LAMBDA>~(x:\<INT>): 0 \leq x)} \\\noalign{\smallskip}
\inference[$\pfun$]{f: X \pfun Y}
  {f:[\#~dom:\<ARRAY>~X~\<OF>~\<BV1>, law:\<ARRAY>~X~\<OF>~Y~\#]} \\\noalign{\smallskip}
\inference[$\emptyset$]{\emptyset: X}
  {emptysetX:\<ARRAY>~X~\<OF>~\<BV1> = (\<ARRAY>~(y:Y): \<0>)} \\\noalign{\smallskip}
\inference[$\cap$]{A,B: \power X & A \cap B}
  {\begin{array}[t]{l}
     capX:(\<ARRAY>~X~\<OF>~\<BV1>, \<ARRAY>~X~\<OF>~\<BV1>) -> \<ARRAY>~X~\<OF>~\<BV1> \\
     \<ASSERT>~\<FORALL>~(A,B:\<ARRAY>~X~\<OF>~\<BV1>): \\
       \t1 capX(A, B) = (\<ARRAY>~(x:X): \<IF>~A[x] = \<1>~\<AND>~B[x] = \<1>~\<THEN>~\<1>~\<ELSE>~\<0>~\<ENDIF>)
   \end{array}
  } \\\noalign{\smallskip}
\inference[$\subseteq$]{A,B: \power X & A \subseteq B}
  {\begin{array}[t]{l}
     subseteqX:(\<ARRAY>~X~\<OF>~\<BV1>, \<ARRAY>~X~\<OF>~\<BV1>) -> \<BOOLEAN> \\
     \<ASSERT>~\<FORALL>~(A,B:\<ARRAY>~X~\<OF>~\<BV1>): \\
       \t1 subseteqINT(A, B) <=> \<FORALL>~(x:\<INT>): A[x] = \<1> => B[x] = \<1>
   \end{array}
  }  \\\noalign{\smallskip}
\inference[$\finset$]{A:\finset X}
  {\begin{array}[t]{l}
     A:[\#~set:\<ARRAY>~X~\<OF>~\<BV1>, bij:\<ARRAY>~X~\<OF>~\<NAT1>, card:\<NAT>~\#] \\
     \<ASSERT>~\<FORALL>~(x:X): A.set[x] = \<1> <=> A.bij[x] \leq A.card \\
     \<ASSERT>~\<FORALL>~(n:\<NAT1>, x_1,x_2:X): \\
       \t1 (n \leq A.card~
           \<AND>~A.set[x_1] = \<1>~
           \<AND>~A.set[x_2] = \<1>~\<AND>~A.bij[x_1] = n~\<AND>~A.bij[x_2] = n) \\ 
       \t1 => x_1 = x_2
   \end{array}
  } \\\noalign{\smallskip}
\inference[$\seq$]{s:\seq X}
  {\begin{array}[t]{l}
     s:[\#~dom:\<ARRAY>~X~\<OF>~\<BV1>, law:\<ARRAY>~\<NAT1>~\<OF>~X, card:\<NAT>~\#] \\
     \<ASSERT>~\<FORALL>~(n:\<NAT1>): n \leq s.card <=> s.dom[n] = \<1>
   \end{array}
  }
\end{array}
\caption{\label{cvc3}Embedding rules for CVC3.}
\end{figure}

\subsection{\label{var}A Variant}
Besides the embeddings shown in Figures \ref{yices1} and \ref{cvc3}, we also tried out a variant for each of them. In this variant the rules for basic types are replaced by the following ones:
\[
\inference[Yices]{[X]}{(\<define-type>~X~(\<scalar>~x_1~x_2~x_3))} \t2 
\inference[CVC3]{[X]}{\<DATATYPE>~X = x_1 | x_2 | x_3~\<END>}
\]

In other words, a basic type is replaced by a type with only three values. Fastest proceeds in a similar fashion as we have explained in Section \ref{genalltca}. Given that the elements of a basic type have an uncertain structure, we have observed that in many test specifications there is no need in having all of them. Changing the rules in this way may have a great impact on the effectiveness of the SMT solvers because all the quantifications over $X$ become finite. It is a known fact that SMT solvers turn out to be incomplete at the presence of quantifications over infinite sets. Therefore, in this way it may be possible to avoid a number of such quantifications thus increasing the likelihood of finding more test cases. 

\section{\label{experiments}Empirical Assessment}
Since we started with the Fastest project we used a number of case studies (Z specifications) to test and validate different aspects of the tool \cite{CristiaPRM, CristiaARM, CSV, CristiaPluss, CristiaSEW2011}. At the moment we have eleven case studies to test
the test case generation algorithm described in Section \ref{genalltca}.
Fastest finds 100\% of the test cases for two of the eleven case studies. Of the remaining nine, we discarded one for the present experiments because it has very long test specifications as to write them by hand. Therefore, we assessed Yices and CVC3 and the embeddings with test specifications from eight case studies. The 69 test specifications chosen for this assessment are those for which Fastest was unable to find a test case, although they are satisfiable. The satisfiability of these test specifications was determined by manual inspection.

All these experiments were conducted on the following platform: an Intel Centrino Duo of 1.66 GHz with 1 Gb of main memory, running Linux Ubuntu 10.04 LTS with kernel 2.6.32-35-generic. As we have said, the original Z test specifications, and their translation to Yices and CVC3 can be downloaded from \verb+http://www.fceia.unr.edu.ar/~mcristia/smt-ttf-cs.tar.gz+. The translation of each test specification is saved in a file ready to be loaded into Yices or CVC3. We also provide scripts to run the experiments. The results can be analysed with simple \verb+grep+ commands.

The first experiment started by manually writing each test specification according to the embedding rules shown in Figures \ref{yices1} and \ref{cvc3}. Then Yices and CVC3 were fed with each of them followed by a check-sat command. The output was redirected to files to be analysed later. The second experiment consisted in applying the variant embedding described in Section \ref{var}. After a check-sat command they both return either ``satisfiable'' or ``unknown''---because ``unsatisfiable'' is impossible as all the test specifications are satisfiable. In both cases ``unknown'' means that the SMT solver cannot decide whether the formula is satisfiable or not. When the answer is ``satisfiable'' both solvers return a witness satisfying the formula. Furthermore, if the answer is ``unknown'' they return a ``potential witness''. That is, they are not sure whether the formula is satisfiable or not but they ``believe'' it is and return a possible witness.

\begin{table}
\begin{tabularx}{\textwidth}{l>{\raggedleft}X>{\raggedleft}X>{\raggedleft}X>{\raggedleft}X>{\raggedleft}X>{\raggedleft}X>{\raggedleft}X>{\raggedleft\arraybackslash}X}\hline
& \multicolumn{4}{c}{\textbf{Embeddings of Figure \ref{yices1} and \ref{cvc3}}} & \multicolumn{4}{c}{\textbf{Variant described in Section \ref{var}}}\\
\multicolumn{1}{c}{\textbf{Case study}} & \multicolumn{2}{c}{\textbf{Yices}} & \multicolumn{2}{c}{\textbf{CVC3}} & \multicolumn{2}{c}{\textbf{Yices}} & \multicolumn{2}{c}{\textbf{CVC3}} \\
& \multicolumn{1}{c}{\textbf{Sat}} & \multicolumn{1}{c}{\textbf{Unk}} & \multicolumn{1}{c}{\textbf{Sat}} & \multicolumn{1}{c}{\textbf{Unk}} & \multicolumn{1}{c}{\textbf{Sat}} & \multicolumn{1}{c}{\textbf{Unk}} & \multicolumn{1}{c}{\textbf{Sat}} & \multicolumn{1}{c}{\textbf{Unk}} \\\hline
Savings accounts (3) & 8 & & 8 & & 8 & & 8 & \\
Savings accounts (1) & & 2 & & 2 & & 2 & & 2 \\
Launching vehicle & & 8 & 8 & & & 8 & 8 & \\
Plavis & & 29 & & 29 & & 29 & & 29 \\
SWPDC & & 13 & & 13 & & 13 & & 13 \\
Scheduler & & 4 & & 4 & & 4 & & 4 \\
Security class & & 4 & & 4 & & 4 & & 4 \\
Pool of sensors & 1 & & 1 & & 1 & & 1 & \\\hline
\textbf{Totals} & 9 & 60 & 17 & 52 & 9 & 60 & 17 & 52 \\\hline
\end{tabularx}
\caption{\label{t:exper}Results of running Yices and CVC3 on 69 test specifications.}
\end{table}

The results of these experiments are shown in Table \ref{t:exper}. Column \textbf{Sat} (\textbf{Unk}) is the number of test specifications for which the SMT solver returned ``satisfiable'' (``unknown''). As it can be seen, the variant embedding produced exactly the same results for both SMT solvers. Also it is easy to see that CVC3 discovered all the test cases discovered by Yices plus eight more. However, while Yices, in both experiments, took no more than 3 seconds in processing the 69 test specifications, CVC3 took around 7 minutes in doing the same. In turn, Fastest takes 6.5 minutes to process the same test specifications, but, as we have already said, it discovers no test case. Yices could not solve all the test specifications that include a quantification or a lambda expression over an infinite set; CVC3 could not solve all the test specifications that include a quantification over an infinite set. It is very important to remark that these quantifications or lambda expressions appear due to the embeddings; they are not present in the original Z test specifications. The conclusions about these experiments are listed in Section \ref{concl}.

\section{\label{relwork}Related Work}
We chose Yices and CVC3 for this work after evaluating all the SMT solvers that participated in the SMT-COMP 2010, 2009 and 2008, that is 22 tools \cite[chapter 5]{Cok00}. The evaluation was based on the following criteria: (i) the tool must be documented as to be used by a novice user, specifically its input language must be thoroughly described; (ii) the tool must be stable and actively developed; (iii) the tool must run on Linux; (iv) the tool must be clearly identified as a SMT solver, specifically it must return the witness satisfying a formula; (v) the tool must be freely available to the general public; and (vi) the tool must work with a general logical system, in particular it must support: (a) quantified formulas; (b) basic and enumerated types; and (c) a general theory of extensional uninterpreted functions or arrays (index and values over general types). This evaluation yielded only three candidates: Yices, CVC3 and Z3 \cite{Bjorner00}. In this paper we report on the results of applying Yices and CVC3; Z3 will be approached soon. Most of the evaluated SMT solvers do not support quantified formulas over a sufficiently general mathematical theory. veriT supports such theories but it does not provide a witness if a formula was satisfied \cite{veriT}; Alt-Ergo looks powerful as to fulfil our needs but it is not documented as to start the project with it \cite{altergo}. None of the evaluated SMT solvers implement decision procedures for a theory of sets.

Model-based testing (MBT) techniques and tools for constraint solving or satisfiability have been integrated, and SMT developers have proposed to use their tools for test case generation. The results of some of these works encouraged us to follow the same ideas but applied to the TTF and Z. For example, Leonardo de Moura, in a tutorial given at Automated Formal Methods (AFM) 2006, includes test case generation as one of the applications for Yices. Different people at Microsoft Research have integrated Z3 into MBT or testing tools. Veanes and colleagues \cite{Veanes01} use Z3 to generate test cases for parametrized SQL queries. In this work the authors use a language which supports finite sets, but not the other ZMT elements. Pex \cite{Pex} and SAGE \cite{SAGE} are testing tools developed at Microsoft Research which integrate Z3. The first one generates unit tests for .NET applications, and the second one is a fuzz tester for security vulnerabilities. Grieskamp et al. \cite{Grieskamp02} use Spec Explorer integrated with Z3 to generate combinations of parameter values. This parameters appear in the actions of labelled transition systems abstracting the system-under-test. Besides, Galler et al. \cite{Galler00} integrate Z3 in jCAMEL so it can derive test cases for programs annotated with contracts. However, Z3 is used only for integer parameters. As it can be seen, these works deal with formalisms quite different from and usually less expressive than Z.

The satisfiability algorithm presented in Section \ref{genalltca} is similar in conception to approaches like the Alloy Analyser which also defines a finite model for a given predicate and tries to see whether is it possible to satisfy it within this model or not \cite{Jackson00}. The Alloy Analyser uses some SAT solving techniques.

Peleska et al. \cite{Peleska01} apply the SONOLAR SMT solver for generating test cases from a modelling formalism based on Harel's Statecharts. This formalism is less expressive than Z and does not include any set or related theory. SONOLAR is one of the SMT solvers we evaluated but we could not use it since it does not support quantified formulas nor a general theory of arrays or uninterpreted functions.

Kr\"oning et al. \cite{Kroning00} propose to add a new theory to the SMT-Lib standard \cite{Barrett01}, as the standard format for formulas involving sets and finite sets, mappings and lists. Their proposal originates in VDM but they acknowledge that it can be applied to other formalisms such as Z. However, they do not give the embedding of this theory in the language of any concrete SMT solver; they suggest that arrays can be used to encode it. Besides, the theory described in that work is not exactly Z---for example, they deal with finite mappings and not functions as in Z.

\section{\label{concl}Conclusions and Future Work}
In this paper we have proposed shallow embeddings for two SMT solvers, Yices and CVC3, as a method for finding test cases from Z test specifications. These test specifications are generated by Fastest, a tool implementing the Test Template Framework. Given that these test specifications are predicates of first-order logic and Zermelo-Fraenkel set theory, SMT solvers looked as promissory tools to solve this problem. Besides, we experimented with these embeddings and the SMT solvers by manually codifying 69 satisfiable test specifications. Based solely on these experimental results we can conclude:

\begin{itemize}
\item CVC3 works better than Yices, as the former found around the double of test cases.
\item Given that CVC3 discovered exactly the same test cases than Yices, combining both tools does not seem to be fruitful, unless Yices is used first given that it runs faster than CVC3.
\item However, CVC3 discovered test cases for only 25\% of the test specifications. It seems a poor result since it outperforms the rough algorithm implemented by Fastest in only 17 test cases. Assuming CVC3 would also discover all the test cases that Fastest currently does---we are not sure it will, though---, it would be an overall increment of 4\%, given that we work with 475 satisfiable test specifications. Furthermore, the time spent by CVC3 and Fastest in procssing these 17 test cases is roughly the same.
\item An issue that deserves more attention is the chances of using the potential witnesses returned by the SMT solver when the answer is ``unknown''. After a manual inspection we observed that many of them are indeed witnesses. It might be possible to invoke ZLive to confirm that these witnesses satisfy their corresponding test specifications. This constitutes a first sign that combining Fastest with SMT solvers may be a good option.
\item The previous item brings in another issue. Is it trivial to automatically translate back to Z the witnesses returned by the SMT solver? At a first glance, the witnesses returned by Yices are far easier to parse than those returned by CVC3---actually Yices returned a total of 1,221 lines of text while CVC3 returned 33,145 lines; the main difference lies in the potential witnesses: less than 1,000 lines for Yices and more than 32,000 lines for CVC3.
\item Replacing the embedding of Z basic types by enumerated types (as described in Section \ref{var}) proved to be useless, in spite of looking promising at first. The problem may lay in the fact that this variant still produces formulas with quantifications over the integers or the naturals---i.e. infinite quantifications. It does not seem promising to change $\num$ or $\nat$ for a finite subset---like, for instance $[-10 \upto 10]$---because each literal has its own properties. For example, if a test specification mentions ``43'' then not considering it in some way may lead to an ``unsatisfiable'' answer. This is not the case for basic types, as all of their elements have only one property: equality.
\end{itemize}

In summary, we will keep exploring combining SMT solvers with Fastest since they discovered some test cases that Fastest did not, their execution times are at least as good as Fastest's, and there are chances that potential witnesses become more test cases. Our next step is to see whether the embedding for CVC3 finds all the test cases that Fastest currently finds and study the translation of the witnesses returned by CVC3. If the results of this step are not good, then we will consider proposing a decision procedure for formulas of the ZMT. We also plan to repeat the work reported here with the Z3 SMT solver.

\bibliographystyle{eptcs}
\bibliography{smt-ttf}
\end{document}